\documentclass[aps,prb,twocolumn,superscriptaddress,showpacs,showkeys,notitlepage]{revtex4-1}
\usepackage{graphicx}
\usepackage{dcolumn}
\usepackage{bm}
\usepackage{float}
\usepackage{amsmath}
\usepackage{xcolor}
\usepackage{textcomp}
\usepackage{lipsum}
\usepackage{upgreek}

\begin{document}

\title{X-ray holographic imaging of magnetic surface spirals in FeGe Lamellae}

\author{L. A. Turnbull}
\email{l.a.turnbull@durham.ac.uk}
\address{Department of Physics, Durham University, Durham, DH1 3LE, UK}

\author{{M. T. Littlehales}}
\address{Department of Physics, Durham University, Durham, DH1 3LE, UK}

\author{M. N. Wilson}
\address{Department of Physics, Durham University, Durham, DH1 3LE, UK}

\author{M. T. Birch}
\address{Max Planck Institute for Intelligent Systems, 70569 Stuttgart, Germany}

\author{H. Popescu}
\address{Synchrotron SOLEIL, Saint Aubin, BP 48, 91192 Gif-sur-Yvette, France}

\author{N. Jaouen}
\address{Synchrotron SOLEIL, Saint Aubin, BP 48, 91192 Gif-sur-Yvette, France}

\author{J. A. T. Verezhak}
\address{Department of Physics, University of Warwick, Coventry, CV4 7AL, UK}

\author{G. Balakrishnan}
\address{Department of Physics, University of Warwick, Coventry, CV4 7AL, UK}

\author{P. D. Hatton}
\address{Department of Physics, Durham University, Durham, DH1 3LE, UK}

\begin{abstract}
Isotropic helimagnets are known to host a diverse range of chiral magnetic states. In 2016, F.N. Rybakov \textit{et al.} theorized the presence of a surface-pinned stacked spin spiral phase [F.N. Rybakov \textit{et al.}, 2016 New J. Phys. 18 045002], which has yet to be observed experimentally. The phase is characterized by surface spiral periods exceeding the host material’s fundamental winding period, $L_{\mathrm{D}}$. Here we present experimental evidence for the observation of this state in lamellae of FeGe using resonant x-ray holographic imaging data and micromagnetic simulations. We find images of FeGe lamellae, exceeding a critical thickness of 300 nm (4.3$L_{\mathrm{D}}$), exhibit contrast modulations with a field-dependent periodicity of $\lambda\geq1.4L_{\mathrm{D}}$,  consistent with theoretical predictions of the stacked spiral state. The identification of this spiral state has significant implications for the stability of other coexisting spin textures, and will help complete our understanding of helimagnetic systems.
\end{abstract}

\maketitle

Broken inversion symmetry in the crystal structure of chiral magnets induces an antisymmetric exchange interaction known as the Dzyaloshinskii-Moriya interaction (DMI) \cite{DZYALOSHINSKY1958,Moriya1960}. 
Competition between the DMI and ferromagnetic exchange interaction in such systems stabilizes a helical ground state, characterized by the incommensurate winding of the magnetization, $\textbf{M}$, about a propagation vector (Fig. 1(a)) \cite{Beille1983}. 
These helimagnetic systems have garnered significant interest due to the rich array of spiral structures arising in their magnetization, including chiral soliton lattices in layered CrNb$_{3}$S$_{6}$ \cite{Togawa2012} and skyrmion lattices in cubic helimagnets such as MnSi \cite{Mulb2009,Yu2015}, CoZnMn alloys \cite{Tokunaga2015,Karube2020} and FeGe \cite{Yu2011,Tang2021}, the material on which this study focuses. 
In particular, the topological and transport properties of these emergent magnetic states show the potential for novel applications in  advanced spintronic devices \cite{Tomasello2014,Zazvorka2019,Song2020,Back_2020}.

The standard model for magnetism in bulk cubic helimagnets takes the form of the energy density functional
\begin{equation}
    w = A(\nabla\cdot\textbf{m})^{2} + D\textbf{m} \cdot(\nabla\times \textbf{m}) - \mathrm{\mu_{0}}M_{s}\textbf{m} \cdot \textbf{H},\end{equation}
where the terms represent the exchange interaction with stiffness constant $A$, DMI with constant $D$ and Zeeman interaction respectively  \cite{Bak1980,Nagaosa2013}. The external magnetic field vector is \textbf{H}, while \textbf{m} is the unit vector in the direction of the local magnetization, $\textbf{M}$=$M_{s}\textbf{m}$. The bulk magnetic phase diagram arising in these archetypal helimagnetic systems is well established: in an increasing external magnetic field the multi-domain helical ground state transforms into a single-domain conical state propagating parallel to the field direction, before ultimately converging to a saturated ferromagnetic state at a critical field of $H_{D} = D^{2}/2AM_{s}$ \cite{Lebech_1989,Buhrandt2013}. A key result is that the equilibrium period of the conical state and the zero-field helicoid state,
\begin{equation}
L_{\mathrm{D}} = \frac{4 \pi A }{|D|},
\end{equation}
is determined by the ratio of the exchange stiffness and DMI constant \cite{Bogdanov1994}. Magnetocrystalline anisotropy terms are commonly neglected in this model, due to their comparatively weak contribution.

Finite thickness effects, such as shape anisotropy and exposed sample boundaries, can also significantly modify the local energy landscape. The energetics of such shape anisotropy effects can be considered in the context of the energy density of the demagnetization field,
\begin{equation}
w_{\mathrm{D}} = -\frac{1}{2}\mathrm{\mu_{0}}M_{s}\textbf{m}\cdot\textbf{H$_{\mathrm{d}}$},
\end{equation}
where \textbf{H$_{\mathrm{d}}$} is the demagnetizing field.
In lamellae of thickness $L$ $\sim$ $L_{\mathrm{D}}$, the magnetization undergoes a field-induced transformation to an equilibrium lattice of skyrmion tubes \cite{Birch2020}, due to the effect of a chiral surface twisting, which lowers the energy of the skyrmion lattice relative to the conical state \cite{Rybakov2013}. In lamellae thicker than the fundamental period,  $L$ $\gtrsim$ $L_{\mathrm{D}}$, stable surface-pinned skyrmions, known as chiral bobbers, which collapse into Bloch points in the bulk of a sample, have also been observed \cite{Zheng2018}. Additionally, it has been demonstrated that the helicity of a skyrmion tube is modified from a Bloch character in the bulk of a sample, towards a N\'eel character at the surface \cite{Zhang2018,Zheng2021}, akin to N\'eel closure caps \cite{Durr1999,Marioni2006}, indicating the complex spiral structures that can emerge at the boundaries of a magnetic system.

In 2016, Rybakov \textit{et al.} theorized the existence of a stacked spin spiral phase (StSS)(Fig. 1(b,c)), which is comprised of surface-pinned spiral modulations which relax toward a bulk conical state embedded in isotropic chiral magnets, for $L$ $\geq$ 4.18$L_{\mathrm{D}}$ \cite{Rybakov_2016}. The phase is characterized by surface modulation periods exceeding the fundamental period, $L_{\mathrm{D}}$, and was predicted to have both Bloch and N\'eel character, akin to the surface helicity of skyrmions. Here we present resonant x-ray holography data and micromagnetic simulations consistent with the observation of this stacked spin spiral state in lamellae of FeGe.
\begin{figure*}
  \includegraphics[width=0.93\textwidth]{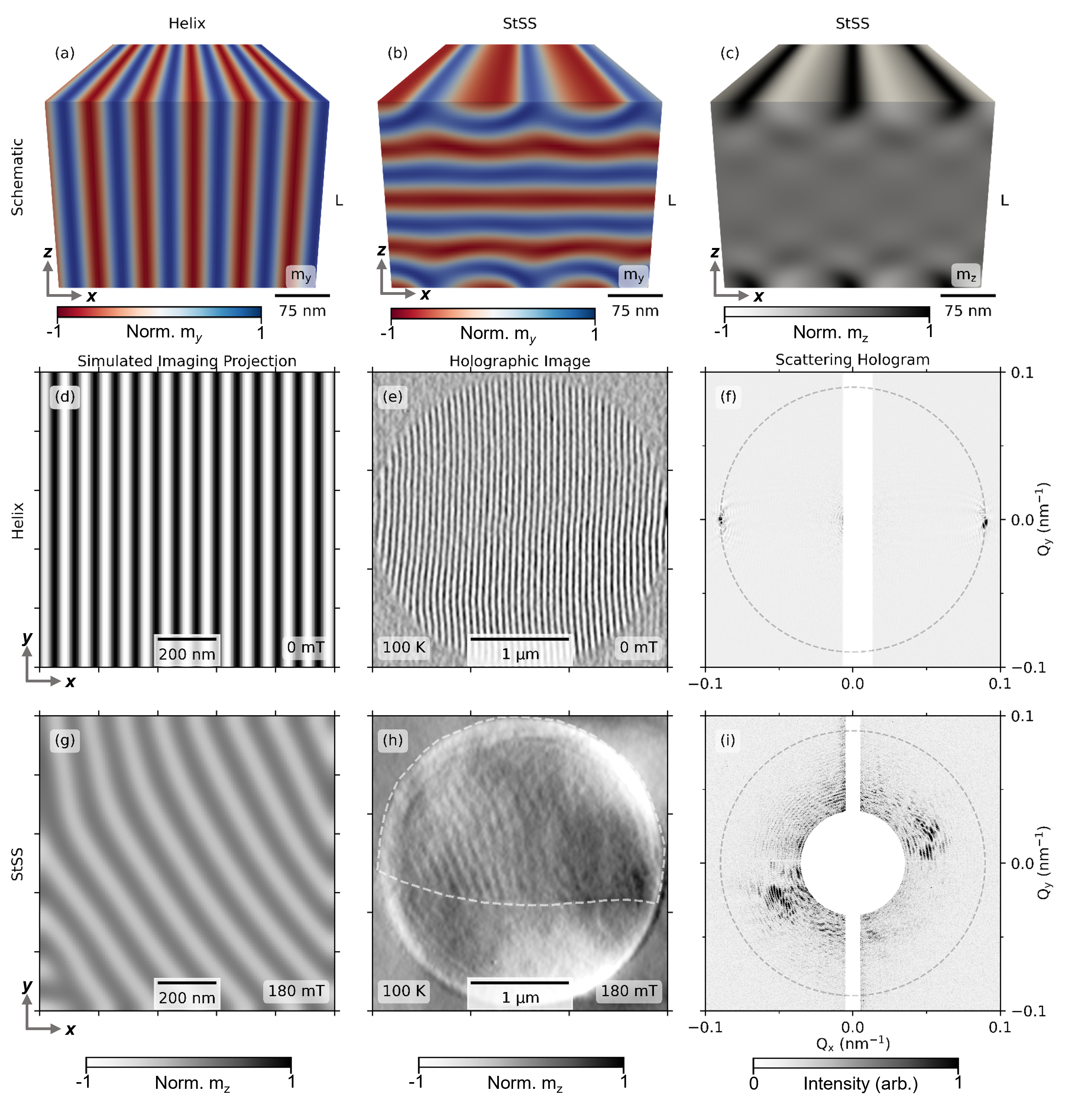}
\caption{a) Schematic of a magnetic helical state, with color map showing the normalized m$_{y}$ component. b) Schematic of a magnetic surface spiral state, with color map showing the normalized m$_{y}$ component. c) Schematic of a magnetic surface spiral state, with color map showing the normalized m$_{z}$ component. d) Simulated x-ray imaging projection of a magnetic helix. Regions of black (white) represent net magnetization towards (away) from the reader. e) X-ray holographic image of a magnetic helix. f) X-ray scattering hologram of a magnetic helix. The dashed circle marks the q-range for the zero-field helical state. The direct transmission of the beam is masked for clarity. g) Simulated x-ray imaging projection of the surface spiral state. h) X-ray holographic image of a surface spiral state. i) X-ray scattering hologram of a magnetic surface spiral. The direct transmission from the holography apertures is masked for clarity.}
\label{fig:1}
\end{figure*}

Resonant x-ray holography is a form of coherent diffractive imaging where the phase information of a magnetic state is encoded in a diffraction pattern by the interference of a reference beam with light scattered in transmission through the sample \cite{Eisebitt2004}. This hologram is a reciprocal space map of the sample, and Fourier transforming the hologram reconstructs a real-space holographic image of the magnetization \cite{Zhu2010}. Here we employ an extended reference slit approach, in order to enhance the resolution of the reconstruction \cite{Guizar-Sicairos:07,Turnbull2020}. The magnetic scattering contrast was resonantly enhanced by tuning the x-ray energy to the $L_{3}$ absorption edge of the magnetic atom (Fe-$L_{3}$ = 706 eV) \cite{Blume1988} and the $m_{z}$ components of the magnetization, parallel with the x-ray beam, were isolated by utilizing x-ray magnetic circular dichroism (XMCD) \cite{VDL2014}. 
\begin{figure}  
\includegraphics[width=0.44\textwidth]{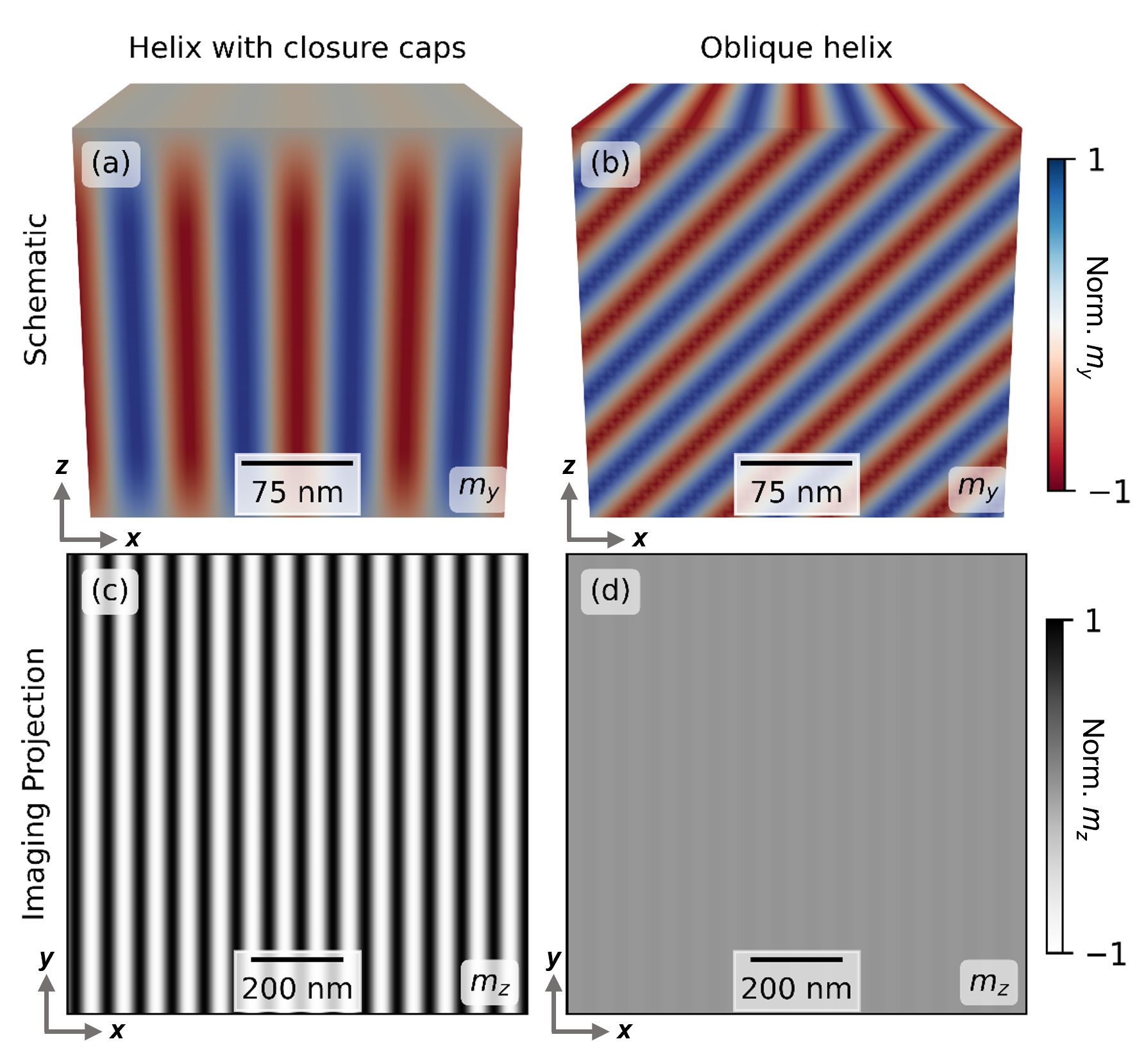}
\caption{a) Schematic of a magnetic helical state exhibiting closure caps at the sample surfaces, with color map showing the normalized m$_{y}$ component. b) Schematic of a magnetic helical state propagating in the [10$\overline{1}$] direction, with color map showing the normalized m$_{y}$ component. c) Simulated x-ray imaging projection of a magnetic helix in panel 2(a). Regions of black (white) represent net magnetization towards (away) from the reader. d) Simulated x-ray imaging projection of a magnetic helix in panel 2(b).}
\label{fig:2}
\end{figure}

\begin{figure}  
\includegraphics[width=0.44\textwidth]{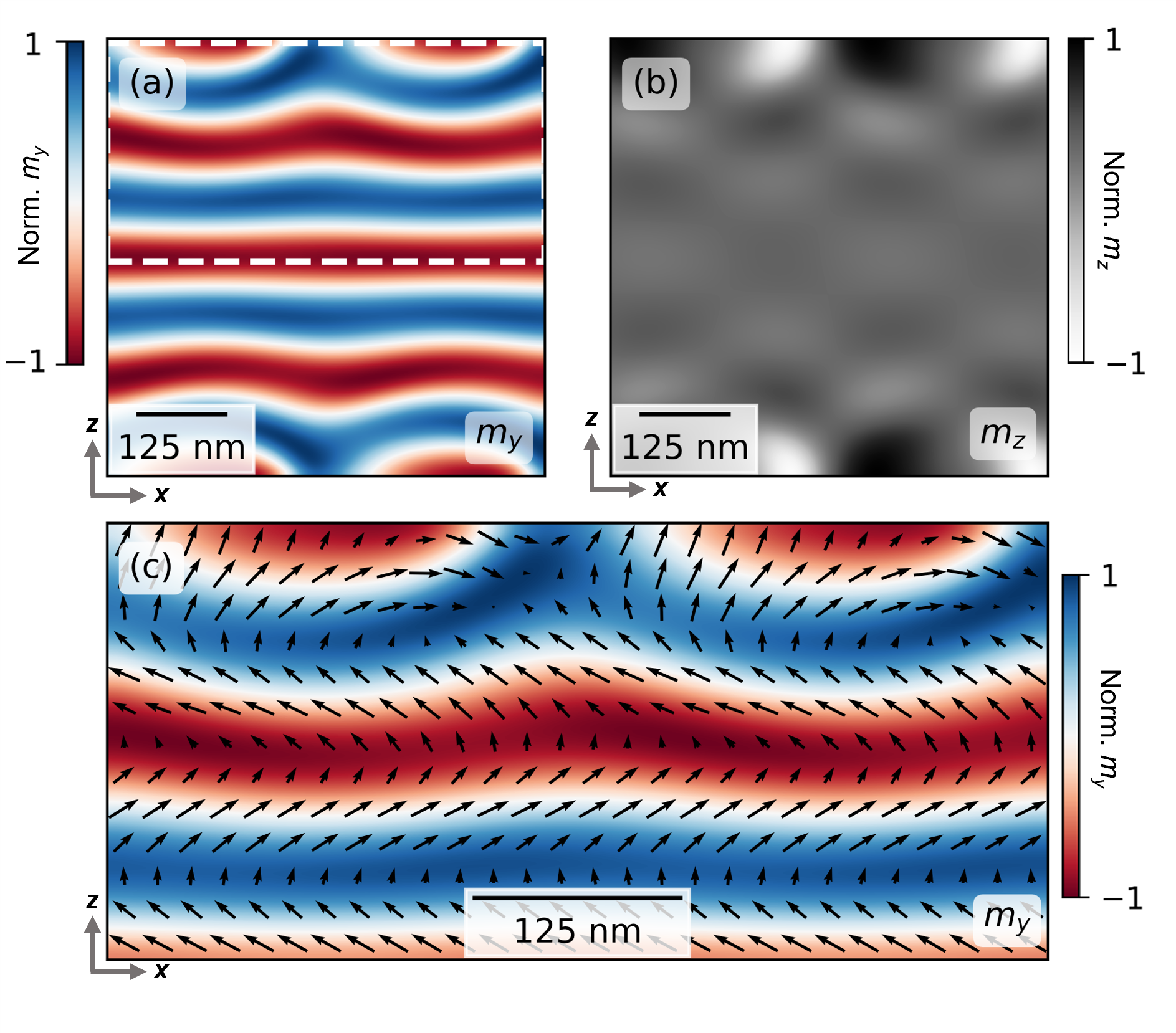}
\caption{Stacked spiral state configuration. a-b) Cross sectional view of the $m_{y}$ and $m_{z}$ components of the magnetization for the surface spiral phase in the $x$-$z$ plane. c) Closer schematic view of the surface spiral, corresponding to the region marked in panel 3(a).}
\label{fig:2}
\end{figure}

\begin{figure*}
  \includegraphics[width=0.95\textwidth]{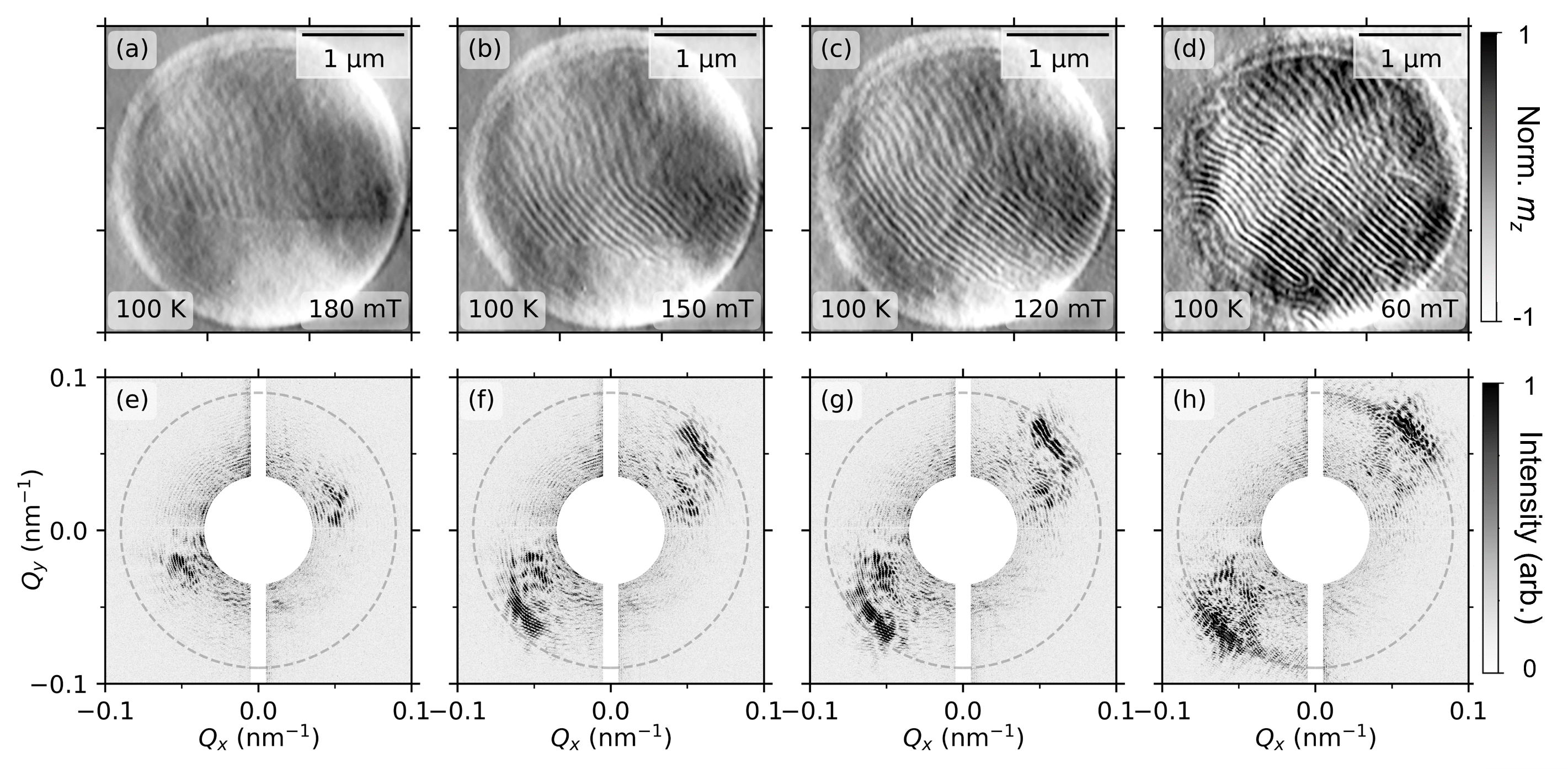}
\caption{Field evolution of the stacked spiral state. a-d) Holographic images of FeGe lamellae with the applied magnetic field marked on each sub-figure. e-f) Corresponding XMCD scattering holograms. The dashed circles marks the q-range for the zero-field helical state. The direct transmission of the beam is masked for clarity.}
\label{fig:3}
\end{figure*}

Figure 1(a) shows the schematic of a single-domain helical state propagating in the $x$-direction and the corresponding simulated x-ray imaging projection along the $z$-direction is shown in Fig. 1(d); regions of white and black represent net magnetization towards and away from the reader. Fig. 1(e,f) show the equivalent experimental x-ray scattering holographic image and hologram of such a helical state in a lamella of FeGe. This sample was 200 nm ($L$=2.85$L_{\mathrm{D}}$) thick and exhibited the expected fundamental winding period of $L_{\mathrm{D}}$ = 70 $\pm$ 2 nm \cite{Burn2019}. The purely sinusoidal modulation of this magnetization structure causes an individual helical domain to comprise a single pair of peaks in reciprocal space. As the relevant forms of the exchange and DM interactions are isotropic, this also defines a fixed radius of 2$\pi$/$L_{\mathrm{D}}$, at which all such peaks would be expected to exist, in the absence of higher order anisotropic terms \cite{Ukleev2021}. The dashed circles in Fig. 1(f,i) mark this observed $q$-value of the zero-field helical state.

Magnetic helices are known to evolve with increasing magnetic field, when pinning due to uniaxial or shape anisotropies resists transition to a conical or field-polarized state, forming a distorted helicoid with an increased period\cite{Izyimov1984,Wilson2020}. In lamellar samples exceeding $L\geq4.3L_{\mathrm{D}}$ in thickness, we observe an extended period modulated state stabilized in out-of-plane magnetic fields, however it does not match the expected distortion of helices in an applied field, and instead shows significant similarities to the surface spiral state predicted by Rybakov \textit{et al}. \cite{Rybakov_2016}.
A holographic image and scattering pattern of such an extended-period modulated state in a 300 nm ($L$=4.3$L_{\mathrm{D}}$) thick  lamella of FeGe are shown in Fig. 1(h,i). The state was produced by zero field cooling from above the Curie temperature ($T_{\mathrm{C}}$  = 278 ± 1  K), to 100 K, increasing the out-of-plane magnetic field to saturation and then decreasing the field at fixed temperature to 180 mT. The top region highlighted in the image shows a series of extended modulations with an ordering period of $(1.83 \pm 0.06)~L_{\mathrm{D}}$ that exhibit 12 \% of the average peak to peak contrast compared to the helical state. The bottom region of the image has no modulated contrast, consistent with an out-of-plane conical state. Figure 1(g) shows the x-ray projection of a simulated surface spiral state stabilized in a 1$\times$1 $\upmu$m$^{2}$ region, which emerged when relaxing a pure out-of-plane conical state, under a 180 mT field aligned with the $z$-axis,
in our micromagnetic simulations \cite{beg2021} using FeGe material parameters \cite{Beg2019}. This projection shows strong qualitative similarities with the holographic image of Fig. 1(h) and exhibits the same ordering behaviour as the scattering hologram.

In order to preclude the observation of a modified archetypal helical state, we simulated the projection of helices with N\'eel closure caps and helices propagating at oblique angles to the sample surface, finding no strong agreement with our observations. 
Figure 2(a,b) show examples of schematic views of such states and the corresponding equivalently normalized imaging projections are visualized in Fig.2(c,d). 
There is minimal modification to the appearance of a magnetic helix with the inclusion of closure caps. The projection through oblique helices causes a reduction in the apparent contract, due to a reduction in the magnetization parallel with the probe beam and a cancelling effect from anti-aligned moments along the $z$-axis. 
While this reduced contrast is in qualitative agreement with the observed state, the oblique helices do not reproduce the extended periodicity of 1.83$L_{\mathrm{D}}$ and are not energetically stable states in the context of the magnetic field conditions and Hamiltonian discussed above. We conclude our observations are likely of the StSS and choose to examine this state in closer detail.

Figure 3 shows cross-sectional views of the $m_{y}$ and $m_{z}$ components in the $x-z$ plane of the simulated surface spiral state. The horizontal stripes in the center of Fig. 3(a) are representative of an out-of-plane conical state, while the top and bottom surfaces show extended period modulations, which decay into the bulk of the sample. Fig. 3(b) shows that the dominant contributions to the x-ray contrast ($m_{z}$), originate at the sample surfaces. In a semi-infinite crystal one would expect the magnetization to transition into a completely pure conical state, however in lamellae of this thickness range the influence of the surface-induced modulation penetrates through the majority of the sample, which can be seen in the weak checker-board pattern occurring throughout Fig 3(b). For a pure conical state propagating in the $z$-direction with a fixed cone angle, one would not expect any modulated contrast in the $m_{z}$ component.

Another notable feature of this surface state is that it does not exhibit a purely helical winding, but rather has a mixed cycloidal and helical winding (as shown in the closer schematic view of Fig. 3(c)) – akin to the helicity of skyrmion tubes changing towards a N\'eel character at the surface \cite{Zhang2018}. The complex spiral structure of this state necessitates higher-order Fourier components than those visible in Fig. 1(h), however their amplitudes are significantly lower than the visible first-order peaks, and one would therefore not expect to readily observe them within our experimental noise floor.
Although it is not possible to fully resolve the chirality or helicity of this state from our experimental imaging geometry alone, future work with tomographic x-ray measurements or complementary measurements from techniques such as LTEM or magnetic force microscopy would be able to do so.

To further investigate this state, Figure 4 shows holographic images and the corresponding scattering holograms of the sample in a continuing downward field sweep from the 180 mT state of Fig 1(h).
When reducing the field, the original region of $(1.83 \pm 0.06)~L_{\mathrm{D}}$ periodicity  remains, however, a state with wavelength $(1.17 \pm 0.03)~L_{\mathrm{D}}$ also emerges in the bottom section of the image.
As the field sweep continues downward, this shorter-wavelength state occupies an increasingly large volume fraction of the sample and its period relaxes towards the fundamental helimagnetic period, while the surface spiral region remains at a higher period of $(1.50 \pm 0.04)~L_{\mathrm{D}}$ at 60 mT. The behaviour in the lower region of the sample is consistent with a common helicoidal state, which is more energetically stable than an out-of-plane conical state in lower magnetic fields. The helicoidal state is also lower in energy than the surface spiral at zero field, except in the limit $L$/$L_{\mathrm{D}}$ $\xrightarrow{}\infty$ \cite{Rybakov_2016}.

Figure 5(a) shows the field dependence of the periodicity of both states. The solid red (purple) line shows the well documented theoretical periods for the helicoid (conical) state \cite{Izyimov1984,Wilson2020}, while the blue line shows the theoretical surface spiral period from Rybakov \textit{et al.} \cite{Rybakov_2016}. This data shows that the field dependence of the two states is significantly different, and that the longer wavelength state matches the predictions of Rybakov \textit{et al}. for the surface spiral. This gives significant additional evidence that this long-wavelength state is the predicted surface spiral state. 

The other notable feature of this helicoidal state is that it increases in contrast relative to the long-period state. As all of the images in Fig. 4 are equivalently normalized, this is indicative of a higher volume fraction of the magnetization aligning parallel with the beam, consistent with the helicoid occupying the full volume of the sample, while the long-period spiral state only generates significant contrast at the surfaces. Surface induced modulations are expected to exhibit an exponential decay into the bulk of a sample over length scales comparable to $L_{\mathrm{D}}$ \cite{Meynell2014}.
The two states also appear to continuously merge into each other, with bifurcation defects mediating the changing wavelength. This could indicate a continuity of striped modulations on the surface, but a difference in how far the underlying states penetrate into the bulk of the sample. It gives the appearance that the helicoids nucleate from the surface state, before growing into the bulk.

\begin{figure}
  \includegraphics[width=0.43\textwidth]{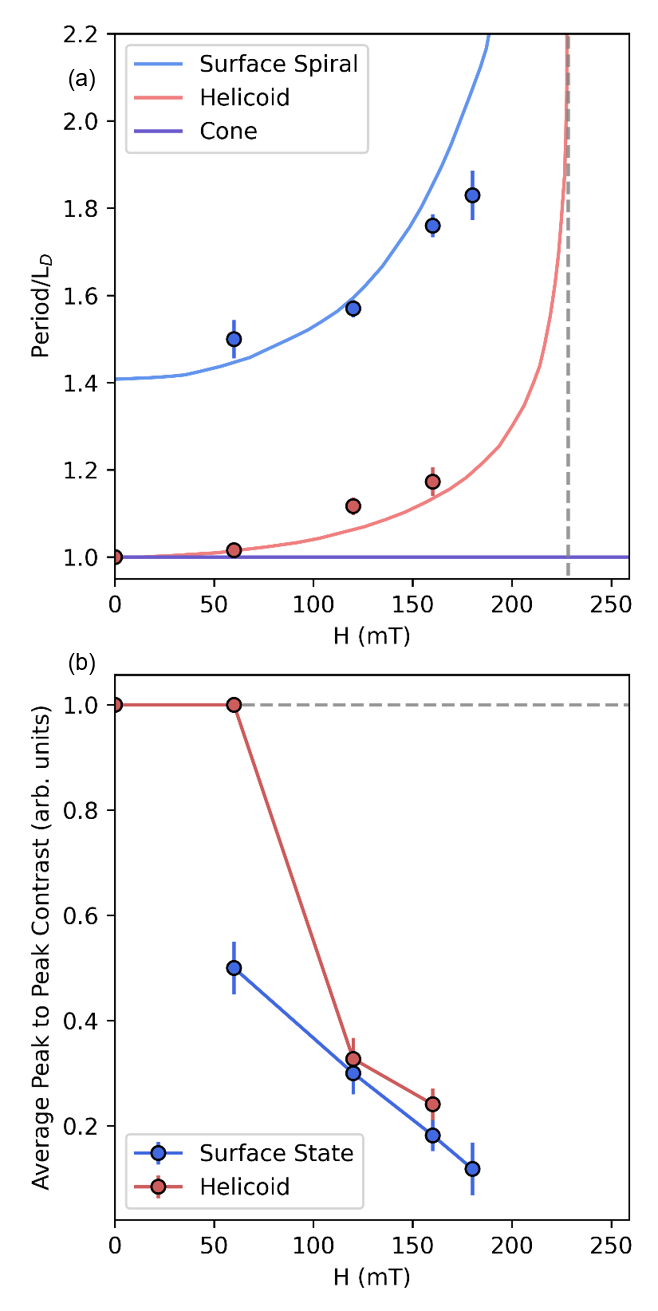}
\caption{Period and contrast of the surface state. a) Periodicity of the surface spiral (blue), helicoids (red) and conical (purple) state as a function of applied field. The solid lines correspond to the well established solutions for the period of helicoids and cones \cite{Izyimov1984,Wilson2020} and the blue line comes from the periodicity established by Rybakov \textit{et al}. b) The equivalently normalized average peak-to-peak contrast of the respective states.}
\label{fig:4}
\end{figure}

Figure 5(b) shows the average peak to peak contrast of both states as a function of field. The surface spiral shows a linear trend, which could indicate either an increasing volume fraction with decreasing field, or an evolution of the fundamental structure of the surface spiral, increasing the surface magnetization parallel with the probe beam.
Notably, the helicoid initially exhibits the same contrast as the surface state at 160 and 120 mT, within error, before ultimately saturating at the higher zero-field signal, while at 60 mT the surface spiral contrast remains low. This is further evidence that the helicoid state could nucleate out from the surface spiral, before penetrating further into the depths of the sample.

To conclude, x-ray holography was used in this study to observe a low-contrast modulated magnetic state with a period exceeding the fundamental helical period of FeGe, that coexists with magnetic helices. These experimental observations demonstrate the existence of the stacked surface spiral state previously proposed in the theoretical predication by Rybakov \textit{et al.} in 2016 \cite{Rybakov_2016}. It likely  that a similar surface state is present in each of the wide array of isotropic chiral magnetic systems currently under investigation. It has been shown that the background state in which skyrmions are embedded modifies their structure and interaction potential. In particular, skyrmions embedded in conically modulated background states are known to exhibit an attractive interaction potential \cite{Leonov_2016}, distinct from the repulsive interaction potential they experience when embedded in a uniformly polarized or helical background state \cite{Du2018}. Therefore, the presence of this spiral surface state will likely modify the behaviour of other spin textures in these materials (such as skyrmions). Depth-dependent studies, such as scattering in reflection geometry or tomographic imaging could be used to map the 3D structure of this state in such future work. 

\begin{acknowledgments}
 We acknowledge SOLEIL for provision of synchrotron radiation facilities and we would like to thank H. Popescu and N. Jaouen for assistance in using beamline SEXTANTS. We acknowledge Diamond Light Source for time on Beamline I10 under Proposal MM27196-1. We acknowledge the GJ Russell Microscopy Facility for provision of focused ion-beam microscopes. This work was supported by the UK Skyrmion Project EPSRC Programme Grant (EP/N032128/1).
\end{acknowledgments}

\bibliography{turnbull2021}

\end{document}


\title{Supplementary Information for Real-space experimental observation of magnetic surface spirals in FeGe}

\author{L. A. Turnbull}
\address{Department of Physics, Durham University, Durham, DH1 3LE, UK}

\author{{M. T. Littlehales}}
\address{Department of Physics, Durham University, Durham, DH1 3LE, UK}

\author{M. N. Wilson}
\address{Department of Physics, Durham University, Durham, DH1 3LE, UK}

\author{M. T. Birch}
\address{Max Planck Institute for Intelligent Systems, 70569 Stuttgart, Germany}

\author{H. Popescu}
\address{Synchrotron SOLEIL, Saint Aubin, BP 48, 91192 Gif-sur-Yvette, France}

\author{N. Jaouen}
\address{Synchrotron SOLEIL, Saint Aubin, BP 48, 91192 Gif-sur-Yvette, France}

\author{J. A. T. Verezhak}
\address{Department of Physics, University of Warwick, Coventry, CV4 7AL, UK}

\author{G. Balakrishnan}
\address{Department of Physics, University of Warwick, Coventry, CV4 7AL, UK}

\author{P. D. Hatton}
\address{Department of Physics, Durham University, Durham, DH1 3LE, UK}

\maketitle

Supplementary figure 1 illustrates the difference arising in the imaging projection due to the inclusion of N\'eel closure caps on a helical state in a sample of FeGe, 280 nm thick, equivalent to the sample used in the main study. It shows there is no qualitative difference in the contrast generated, primarily due to the fixed winding length imposed by the competition of the exchange and Dzyaloshinskii Moriya interactions. Supplementary figure 2 illustrates the difference in the imaging projection due of a magnetic helix with N\'eel closure caps and the stacked spin spiral state reported in the main manuscript. Both states are shown in a 280 nm sample of FeGe and are displayed on an equivalent scale. There is a clear difference in the periodicity and contrast observed. This shows that our data is inconsistent with the presence of just N\'eel closure caps, and requires the addition of the surface state.

\begin{figure}
\includegraphics[width=0.44\textwidth]{SI1.png}
\caption{Comparison of the imaging projection of magnetic helices with and without N\'eel closure caps. a-b) Schematic view of magnetic helices with and without N\'eel closure caps respectively. c-d) Cross sectional view of the $m_{y}$ and $m_{z}$ components of the magnetization for the respective helical states in the $x$-$z$ plane. e-f) X-ray imaging projection of the respective states along the $z$-axis.}
\end{figure}

\begin{figure}  
\includegraphics[width=0.44\textwidth]{SI2.png}
\caption{Comparison of the imaging projection of magnetic helices with N\'eel closure caps and the stacked spin spiral state. a-b) Schematic view of magnetic helices with N\'eel closure caps and the stacked spin spiral state respectively. c-d) Cross sectional view of the $m_{y}$ and $m_{z}$ components of the magnetization for the respective helical and spiral states in the $x$-$z$ plane. e-f) X-ray imaging projection of the respective states along the $z$-axis.}
\end{figure}

\begin{figure}  
\includegraphics[width=0.44\textwidth]{SI3.png}
\caption{a) Schematic of a magnetic helical state, with color map showing the normalized m$_{y}$ component. b) Schematic of a magnetic surface spiral state, with color map showing the normalized m$_{y}$ component. c) Schematic of a magnetic surface spiral state, with color map showing the normalized m$_{z}$ component.}
\end{figure}